\documentclass[twocolumn]{aastex631}
\usepackage{amsmath}	
\usepackage{amssymb}	
\usepackage{xcolor}

\newcommand{\be}{\begin{equation}}
\newcommand{\ee}{\end{equation}}
\newcommand{\units}[1]{\ensuremath{\, \mathrm{#1}}}  
\newcommand{\Msun}{{\rm M}_\odot}
\newcommand{\Mjup}{{\rm M}_{\rm Jup}}

\newcommand{\Lsun}{{\rm L}_\odot}

\DeclareMathOperator\erf{erf}

\begin{document}

\title{Leaning Sideways: VHS~1256$-$1257 b is a Super-Jupiter with a Uranus-like Obliquity}

\correspondingauthor{Michael Poon}
\email{michael.poon@astro.utoronto.ca}

\author[0000-0001-7739-9767]{Michael Poon}
\affiliation{Department of Astronomy and Astrophysics, University of Toronto, Toronto, Ontario, M5S 3H4, Canada}

\author[0000-0002-6076-5967]{Marta L. Bryan}
\affiliation{Department of Chemical and Physical Sciences, University of Toronto Mississauga, Mississauga, Ontario, L5L 1C6, Canada}
\affiliation{Department of Astronomy and Astrophysics, University of Toronto, Toronto, Ontario, M5S 3H4, Canada}

\author[0000-0003-1927-731X]{Hanno Rein}
\affiliation{Department of Physical and Environmental Sciences, University of Toronto Scarborough, Toronto, Ontario, M1C 1A4, Canada}
\affiliation{Department of Astronomy and Astrophysics, University of Toronto, Toronto, Ontario, M5S 3H4, Canada}

\author[0000-0002-4404-0456]{Caroline V. Morley}
\affiliation{Department of Astronomy, The University of Texas at Austin, Austin, TX 78712, USA}

\author[0000-0001-7875-6391]{Gregory Mace}
\affiliation{Department of Astronomy, The University of Texas at Austin, Austin, TX 78712, USA}

\author[0000-0003-2969-6040]{Yifan Zhou}
\affiliation{Department of Astronomy, University of Virginia, Charlottesville, 22904, USA}

\author[0000-0003-2649-2288]{Brendan P. Bowler}
\affiliation{Department of Astronomy, The University of Texas at Austin, Austin, TX 78712, USA}

\begin{abstract}

We constrain the angular momentum architecture of VHS~J125601.92-125723.9, a 140 $\pm$ 20 Myr old hierarchical triple system composed of a low-mass binary and a widely-separated planetary-mass companion VHS~1256~b. VHS~1256~b has been a prime target for multiple characterization efforts, revealing the highest measured substellar photometric variability to date and the presence of silicate clouds and disequilibrium chemistry. Here we add a key piece to the characterization of this super-Jupiter on a Tatooine-like orbit; we measure its spin-axis tilt relative to its orbit, i.e. the obliquity of VHS 1256 b. We accomplish this by combining three measurements. We find a projected rotation rate $v \sin{i_p} = 8.7 \pm 0.1 \units{km~s^{-1}}$ for VHS~1256~b using near-IR high-resolution spectra from Gemini/IGRINS.  Combining this with a published photometric rotation period indicates that the companion is viewed edge-on, with a line-of-sight spin axis inclination of $i_{\rm p} = 90^\circ \pm 18^\circ$. We refit available astrometry measurements to confirm an orbital inclination of $i_{\rm o} = 23 \substack{+10 \\ -13}^\circ$. Taken together, VHS~1256~b has a large planetary obliquity of $\psi = 90^\circ \pm 25^\circ$. In total, we have three measured angular momentum vectors for the system: the binary orbit normal, companion orbit normal, and companion spin axis. All three are misaligned with respect to each other. Although VHS~1256~b is tilted like Uranus, their origins are distinct. We rule out planet-like scenarios including collisions and spin-orbit resonances, and suggest that top-down formation via core/filament fragmentation is promising.

\end{abstract}


\section{Introduction} \label{sec:intro}

Planetary obliquities, the orientation between a planet's spin axis and its orbit normal, are a new window into the formation and evolutionary histories of exoplanets. Up until 2020, only our Solar System planets had measured obliquities. For instance, Uranus rotates on its side, Venus spins upside-down, and Saturn is tilted by 27 degrees, pointing to histories of processes like giant impacts and secular spin-orbit resonances \citep{Ward+Hamilton2004, Correia2006, Nesvorny2018, Reinhardt+2020, Lu+Laughlin2022}. 

Like our Solar System planets, exoplanets could also exhibit a similar diversity of obliquities. In addition to processes suggested for our Solar System planets, theoretical work suggests that planet-disk interactions \citep{Millholland+Batygin2019, Su+Lai2020, Martin+2021}, mergers \citep{Li+Lai2020}, stellar flybys \citep{Rodet+Lai2022}, planet-planet scattering \citep{Li2021}, and disk instability \citep{Jennings+Chiang2021} can also excite planet obliquities. 

\begin{figure*}
    \centering
    \includegraphics[width=0.65\linewidth]{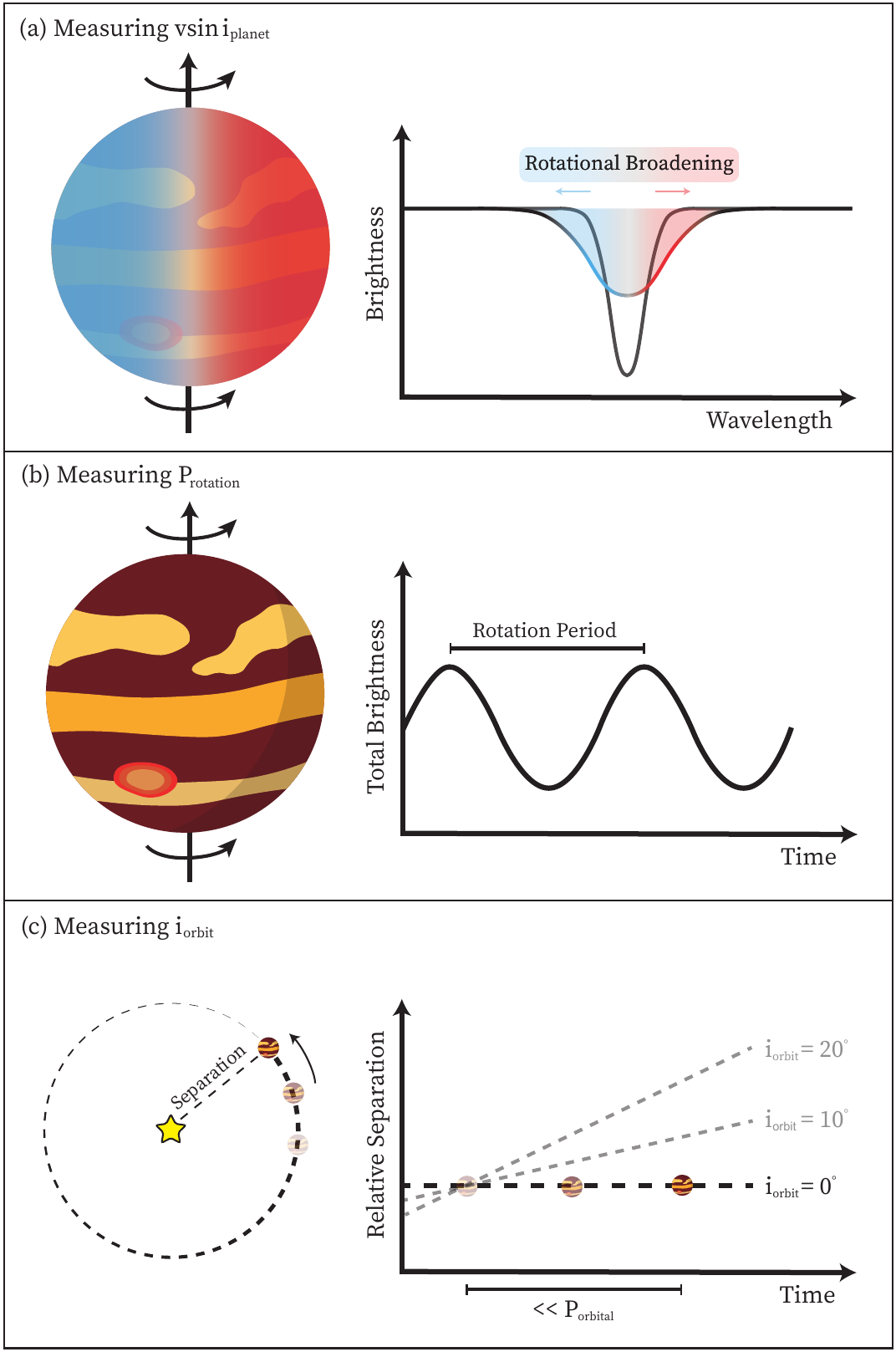}
    \caption{Three observables are required to constrain a planet obliquity: \textbf{(a)} $v \sin{i_\text{planet}}$: High resolution spectroscopy can resolve individual spectral lines to probe rotational line broadening, which yields a projected equatorial velocity ($v \sin{i_\text{planet}}$) \citep{Bryan+2018,Bryan+2020,Bryan+2021}. \textbf{(b) P$_{\rm rot}$:} Time-resolved photometry can monitor variability due to cloud patchiness or longitudinal bands, similar to those seen on Jupiter \citep{Zhou+2020,Zhou+2022}, in order to constrain the rotation period. \textbf{(a)} and \textbf{(b)} can be combined to constrain the planet spin-axis inclination ($i_\text{planet}$) along our line of sight. \textbf{(c) $i_{\rm orbit}$:} Astrometric measurements over a sufficiently long baseline can resolve orbital motion and allow fits to constrain the orbital inclination (\citealt{Bryan+2020,Nguyen+2021,Dupuy+2023}, and references therein).
    }
    \label{fig:measuring_obliquities}
\end{figure*}

Today there are three published planetary obliquities for 2M0122-2439~b, HD~106906~b, and AB~Pic~b \citep{Bryan+2020, Bryan+2021, Palma-Bifani+2023}\footnote{
    We note that this is distinct from the widely studied `stellar obliquity', which instead measures the orientation between the \textit{star's} spin axis and planet orbit normal (see the review by \citealt{Albrecht+2022}). There are stellar obliquity measurements for $\sim 100$ systems, and only currently three exoplanetary obliquity measurements. 
}. These are all young, planetary-mass objects orbiting very far from their host stars ($\sim$ 50-1000 AU). These biases reflect the observational challenges of measuring a planetary obliquity. To do so three observables are required: a projected equatorial velocity ($v \sin{i_p}$, where $i_p$ is the planet spin axis inclination relative to our line-of-sight) from high-resolution spectra of the companion itself, a rotation period from time-series photometry, and an orbital inclination from astrometric measurements. The projected velocity and rotation period require the companion to be bright, favoring young, hot, massive objects far from their brighter host stars. However, if the companions are too far out, constraining their orbits is not feasible. We describe how these measurements are made in Figure~\ref{fig:measuring_obliquities}.

Currently, the population of directly-imaged planetary-mass companions are the only objects amenable to this measurement \citep{Bowler+2016}. We use the term ``planetary-mass companion" to describe these objects that fall into a mass range where bottom-up formation and top-down formation overlap, forming either massive ``super-Jupiters'', or low-mass brown dwarfs. We use the term `planet obliquity' instead of planetary-mass companion obliquities for brevity.

Here we present the fourth exoplanetary obliquity measurement. VHS~J125601.92-125723.9~b (hereafter VHS~1256~b) is a planetary-mass companion orbiting a low-mass binary (total mass $\sim 0.141 \pm 0.008\units{\Msun}$, \citealt{Dupuy+2023}) in a hierarchical triple system \citep{Gauza+2015, Rich+2016, Stone+2016}, with relevant system parameters highlighted in Table \ref{tab:VHS1256_properties}. This companion has a bimodal mass posterior peaking at $12 \pm 0.1$ and $16 \pm 1 \units{\Mjup}$, for deuterium-inert/deuterium-fusing evolutionary models respectively, and is on an eccentric ($e = 0.7 \pm 0.1$) orbit $\sim 400\units{au}$ away from the equal-mass host binary VHS~1256~AB (semi-major axis of $1.96 \pm 0.03 \units{au}$) \citep{Dupuy+2023}. Recently, this system has been a prime target for atmospheric characterization -- it is the only planetary-mass companion targeted with spectroscopy by the JWST High Contrast ERS Program \citep{Hinkley+2022}, which obtained a $1-20$ $\mu$m spectrum indicating the presence of silicate clouds and disequilibrium chemistry \citep{Miles+2023}. In addition, extensive photometric monitoring has shown VHS~1256~b to be the most variable substellar object known to date \citep{Bowler+2020b, Zhou+2020, Zhou+2022}. VHS~1256~b has a previously published photometric rotation period and orbital inclination \citep{Zhou+2020,Dupuy+2023}.

\begin{deluxetable}{ll}
\tablecaption{Measured properties of the VHS~1256 system}
\tablehead{
    \colhead{Property} & \colhead{Measurement}
    }
\startdata 
    VHS~1256~AB mass & $0.141 \pm 0.008 \units{\Msun}$ \\
    VHS~1256~AB separation & $1.96 \pm 0.03 \units{au}$ \\
    VHS~1256~AB eccentricity & $0.883 \pm 0.003$ \\
    VHS~1256~b mass & $12.0 \pm 0.1 \units{\Mjup}$ or $16 \pm 1 \units{\Mjup}$ \\
    VHS~1256~b separation & $350 \substack{+110 \\ -150} \units{au}$\\
    VHS~1256~b eccentricity & $0.7 \pm 0.1$ \\
    System age & $140 \pm 20 \units{Myr}$ \\
\enddata
\tablecomments{These measurements are detailed in \citet{Dupuy+2023}. A consistent measurement of companion mass is also in \citet{Miles+2023}.}
\label{tab:VHS1256_properties}
\end{deluxetable}

In this paper we provide the final ingredient, $v \sin{i_p}$. In Section~\ref{sec:observations} we describe new high-resolution spectroscopic observations with IGRINS/Gemini. Section~\ref{sec:analysis} details our subsequent $v \sin{i_p}$ measurement from these spectra, and our constraints on the line-of-sight inclinations for the companion spin axis, the companion orbit normal, and the host binary orbit normal. We then determine the true 3D angles between each pair of angular momentum vectors (planet spin, planet orbit normal, binary orbit normal). With these constraints, we assess a range of formation and evolutionary scenarios in Section~\ref{sec:dicussion}. In particular, we present a potential formation and evolutionary history for this system, distinguishing between planet-like and star-like scenarios. We present our conclusions in Section~\ref{sec:conclusions}.

\section{Observations} \label{sec:observations}

VHS~1256~b was observed on the nights of 2020 February 02 and 05 UT with the Immersion Grating Infrared Spectrometer \citep[IGRINS;][]{IGRINS1,IGRINS2} on the Gemini South telescope. On the first night of observation, four 800 second exposures were taken. Another four exposures of 1347 seconds were taken on the second night. Each observing sequence was followed by an observation of a telluric standard (HIP~67139) at similar airmass.

The single-exposure spectra were extracted using the IGRINS Pipeline Package \citep[PLP;][]{PLP} on AB nodded pairs and provide sky-subtracted and rectified 2D spectra. Each exposure was individually extracted from the 2D PLP outputs. The individual spectra (eight in total) were then telluric corrected by aligning a telluric feature in the spectrum of VHS~1256~b to the matching line in HIP~67139 before dividing each science spectrum by the Vega corrected telluric standard. The wavelength solution was derived in the standard PLP process of matching OH emission lines and then improving the solution using telluric absorption in the telluric standard.

\section{Analysis} \label{sec:analysis}

\subsection{Measuring \texorpdfstring{$v \sin{i_p}$}{vsini} for VHS~1256~b} \label{sec:vsini}

We infer the projected rotation rate $v \sin{i_p}$ of VHS~1256~b by measuring the amount of spectral line broadening due to planet rotation in these high-resolution spectra. In this section, we describe our methodology, which is similar to \citet{Bryan+2018,Bryan+2020,Bryan+2021}.

While reduced spectra were produced across the wavelength range $\sim$1.45--2.52$\units{\mu m}$, we only consider the $K$-band spectra (1.85--2.52$\units{\mu m}$) in subsequent analyses given the low signal-to-noise ratio (SNR) of the $H$-band spectrum. This yields 8 exposures of wavelength-calibrated and telluric corrected spectra each spanning 26 orders. We also remove orders with low signal-to-noise ratio (order-averaged SNR $<$ 5), and mask artifacts from imperfect strong sky line removal that manifested as spikes in the data. We combine the 8 exposures using an error-weighted average.

We measure rotational line broadening in this spectrum using the cross-correlation methodology outlined in \citealt{Bryan+2021}, briefly summarized here. First, we cross-correlate the spectrum with a model atmosphere broadened to the instrumental resolution, yielding the following ``data'' cross-correlation function (CCF):

\be
\text{CCF}(w) = \cfrac{ \sum\limits_{i=1}^n \Big[ d(\lambda_i) \cdot m(\lambda_i - w) \Big] }{
                       \sqrt{\sum\limits_{i=1}^n \Big[d(\lambda_i)\Big]^2} \cdot
                       \sqrt{\sum\limits_{i=1}^n \Big[m(\lambda_i - w)\Big]^2} },
\label{eq:CCF}
\ee

\noindent where $d$ is the data, the flux from the observed spectrum at each wavelength $\lambda_i$, and $m$ is the model spectrum that has been shifted by a wavelength displacement $w$.

For the model spectra, we use atmospheric models from the Sonora model grid \citep{Sonora_model1, Sonora_model2, Morley+2024}. The Sonora models are calculated using the \texttt{EGP} code, and they assume radiative--convective equilibrium and chemical equilibrium. These are described in more detail in earlier works \citep[e.g.,][]{Marley99b, Saumon08, Morley12, Sonora_model2}. The models are post-processed at the high spectral resolution needed here following the approach of \citet{Morley15}. These are custom versions of the `Sonora Diamondback' models \citep{Morley+2024}, which include silicate (Mg$_2$SiO$_4$, MgSiO$_3$), iron, and corundum (Al$_2$O$_3$) clouds; these clouds assume a low sedimentation efficiency ($f_{\rm sed}=0.5$) leading to lofted clouds \citep{Ackerman+Marley2002}. The Sonora models generated for VHS~1256~b all assume a solar metallicity. 

For our initial model, we assumed an effective temperature $T_\text{eff}=1100\units{K}$ and a surface gravity of $\log{(g)}=4.5$, from medium resolution (R $\sim$ 1000--3700) measurements in \citet{Miles+2023}, which used JWST's NIRSpec IFU and MIRI MRS modes for coverage from 1 to 20$\units{\mu m}$. However, in the course of testing how robust our $v \sin{i_p}$ measurement was to model assumptions (detailed below), we found significant discrepancies when we lowered the effective temperature. 
We found that lower temperature models ($T_\text{eff}=1000\units{K}$) would lead to $v \sin{i_p}$ constraints that differ by $>7\sigma$ compared to higher temperature models ($T_\text{eff}=1100-1300\units{K}$, i.e. Table \ref{tab:vsini_model_tests}). This is because at lower temperatures, methane features dominate over carbon monoxide \citep{Lodders+Fegley2002}, and therefore incorrect spectral lines were being broadened. In the subsequent analysis, we use $T_\text{eff}=1200\units{K}$ and $\log{(g)}=4.5$ as our `best-fit' and fiducial model, since models with $T_\text{eff}=1100\units{K}$ and $1300\units{K}$ yielded consistent $v \sin{i_p}$ constraints. These model parameters are broadly consistent with recent results by \citet{Dupuy+2023} and \citet{Petrus+2024} for VHS~1256~b.

In addition to the data CCF, we calculate a ``model'' CCF which compares a model spectrum that has been broadened to the instrumental resolution to the same model additionally broadened by a rotation rate. To implement rotational broadening, we use a direct integration algorithm described by \citet{fast_rot_broad}, which accounts for the wavelength-dependent effects of Doppler shift much faster than traditional convolution methods.

We compare the data CCF to model CCFs in a Bayesian framework using MCMC to simultaneously infer rotational broadening ($v \sin{i_p}$), radial velocity offset, and the instrumental resolution. We use uniform priors on $v \sin{i_p}$ and the radial velocity offset. We seek an informative prior for the instrumental resolution since this effect is degenerate with rotational broadening. To independently measure the effect of instrumental resolution on our data, we use the \texttt{molecfit} routine, which simultaneously fits a telluric model and an instrumental profile defined by a single Gaussian kernel, to the telluric standard (HIP 67139) spectrum \citep{molecfit1, molecfit2}. We leave out 3 orders (2.01--2.08$\units{\mu m}$) where strong absorption features prevent a good continuum fit. We find an instrumental resolution of $R=49300 \pm 4600$, which we use as a Gaussian prior. We also check if the resolution changes significantly within an order, or between the two observing nights. We find that these subsets of instrumental resolution measurements are mutually consistent with each other and with the global resolution measurement to $<2\sigma$. 

\begin{figure*}
    \centering
    \includegraphics[width=1.0\linewidth]{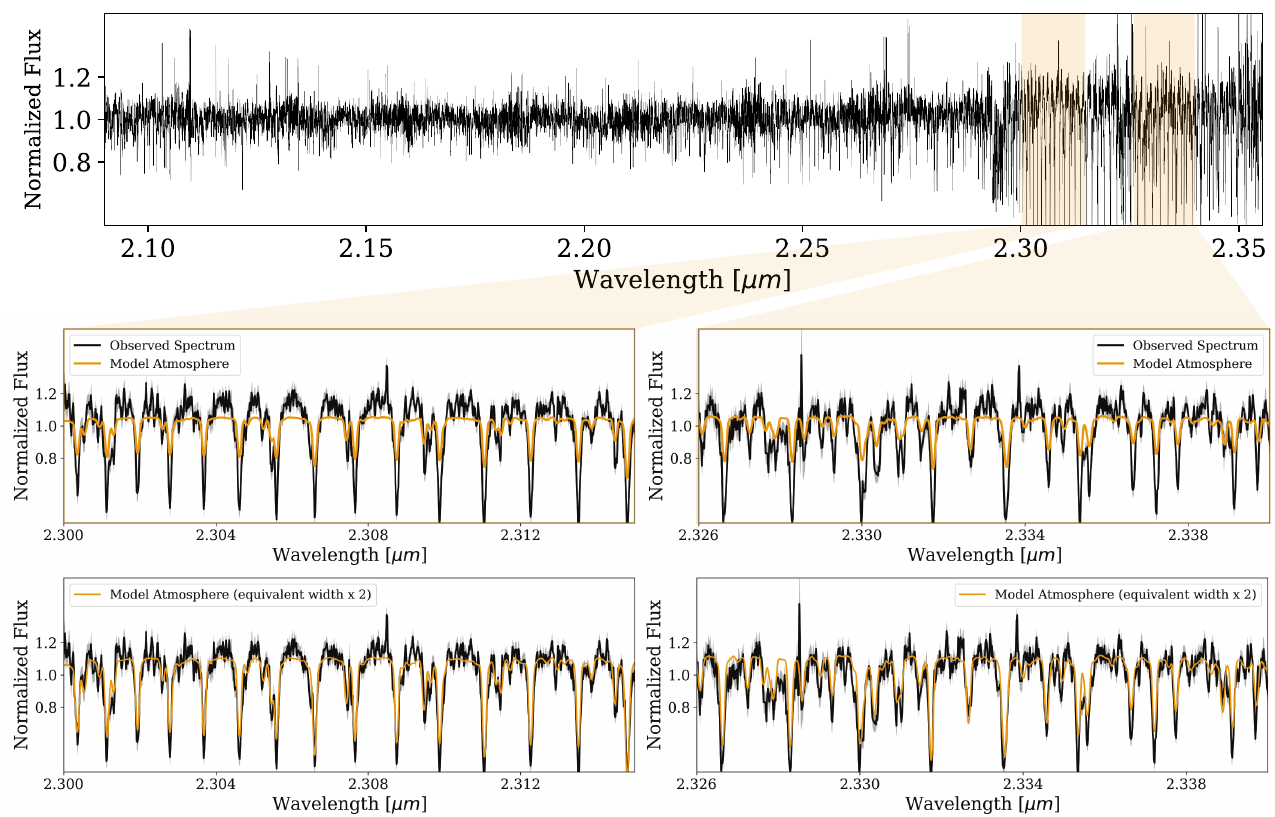}
    \caption{Top: Observed spectrum of VHS~1256~b over wavelength range with higher SNR ($\gtrsim 20$), with portions of order 6 (left) and 5 (right) highlighted. Middle: Zoomed-in panels with $1\sigma$ errors overplotted in grey, and the fiducial model atmosphere broadened to the best-fit projected equatorial velocity overplotted in orange. Although the model atmosphere underestimates the absorption line depths, this does not affect our rotational broadening measurement. We discuss this further in Section \ref{sec:vsini}. Bottom: Same as middle panels, but additionally doubling the equivalent width of the model atmosphere line profiles. We find model assumptions contributing to line-depth mismatch do not significantly impact our measured vsini. 
    }
    \label{fig:spectra}
\end{figure*}

\begin{figure}
    \centering
    \includegraphics[width=1.0\linewidth]{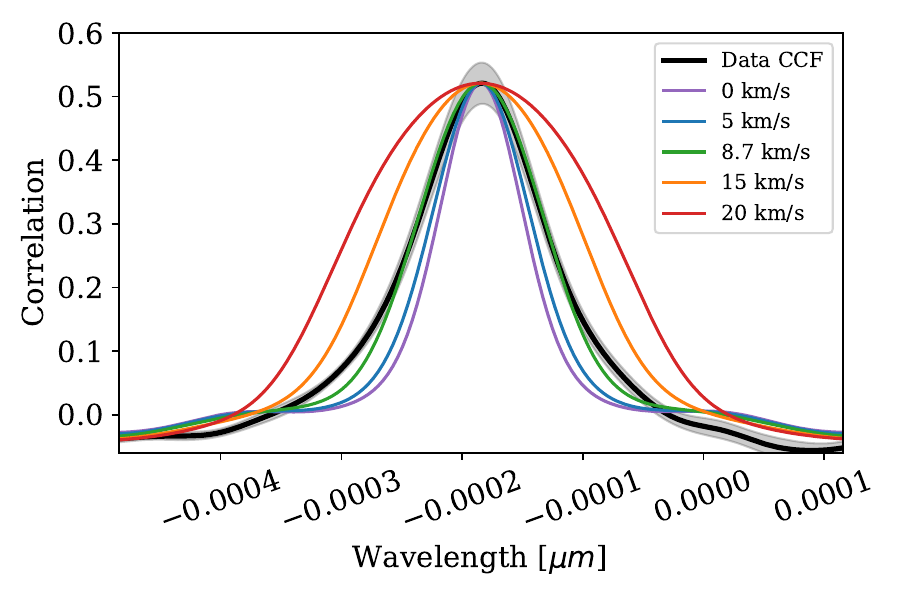}
    \caption{Data CCF (in black) of the observed spectrum cross-correlated with the model atmosphere broadened to the instrumental resolution, with $1\sigma$ jackknife errors shaded in grey. Model CCFs (in color) of the model atmosphere broadened to the instrumental resolution cross-correlated with that same model additionally broadened by a series of projected equatorial velocities (0, 5, 8.7 (best-fit), 15, 20$\units{km~s^{-1}}$).
    }
    \label{fig:CCF}
\end{figure}

The log-likelihood function for our MCMC setup is given by:

\be
\text{log } L = -\frac{1}{2} \sum\limits_{i=1}^n \left[ \left(\frac{M_i - D_i}{\sigma_i}\right)^2 + \ln{(2\pi\sigma^2_i)} \right],
\ee

\noindent where $M$ is the model CCF, $D$ is the data CCF, $n$ is the length of the CCF arrays, and $\sigma$ is the jackknife error given by:

\be
\sigma^2_\text{jackknife} = \frac{(n-1)}{n} \sum\limits_{i=1}^n \big(x_i - x\big)^2,
\ee

\noindent where $n=8$ is the total number of exposures, $x$ is the data CCF using combined observed spectrum, and $x_i$ is the data CCF using the combined observed spectrum excluding exposure $i$. We use a Markov chain Monte Carlo (MCMC) package \texttt{emcee} by \cite{emcee}, and find a projected equatorial velocity $v \sin{i_p}$ of $8.7 \pm 0.1 \units{km~s^{-1}}$ (see Figures~\ref{fig:spectra} \& \ref{fig:CCF} for reference). 

In the middle panel of Figure~\ref{fig:spectra}, we show that the model atmosphere underestimates observed absorption line depths.  This could be due to atmospheric model assumptions that overestimate the thickness of clouds, and/or underestimate the metallicity. \citet{Landman+2023} similarly measured rotational broadening for $\beta$~Pictoris~b with high-resolution spectra in the $K$-band, and find a degeneracy between the effects of clouds and metallicity on absorption line depths. Despite this effect, \citet{Landman+2023} also find that these parameters minimally affect the measured $v \sin{i_p}$ (see their Fig. 4). In the bottom panel of Figure~\ref{fig:spectra}, we show how our fiducial model broadened at our best-fit $v \sin{i_p}$ closely matches the observed line widths when the equivalent width of the model absorption line depths are doubled. We continue with the assumption that our $v \sin{i_p}$ measurement is not significantly affected by the mismatch of absorption line depths.

Next, we test how different assumptions for our atmospheric models could affect the measured $v \sin{i_p}$. Compared to our fiducial model of $T_\text{eff}=1200\units{K}$ and $\log{(g)}=4.5$, we test how sensitive our $v \sin{i_p}$ constraint is to uncertainties of $(100\units{K}, 0.5)$ for effective temperature and surface gravity respectively. These conservative uncertainties are motivated by previous works \citep{Bryan+2020}. Therefore, we generate four new atmospheric models: (1100\units{K}, $\log{(g)}=4.0$), (1100\units{K}, $\log{(g)}=5.0$), (1300\units{K}, $\log{(g)}=4.0$), (1300\units{K}, $\log{(g)}=5.0$) with effective temperatures and surface gravities offset by 1$\sigma$ uncertainties from our fiducial model. We calculate new $v \sin{i_p}$'s with each of these models and find that they are consistent with our original measurement at the $\leq 1.6\sigma$ level (see Table \ref{tab:vsini_model_tests}).

In addition, we test our solar (0.54) C/O assumption by generating new models with 50\% and 150\% solar C/O. We calculate new $v \sin{i_p}$'s with each of these models and find that they are consistent with our original measurement at the $\leq 0.7\sigma$ level (see Table \ref{tab:vsini_model_tests}).

Finally, to test our pressure broadening assumptions, we run two test models with modified molecular cross-sections to simulate $0.1\times$ and $10\times$ the actual pressure for the whole profile. These modified cross-sections would represent extreme uncertainties in how molecular features are pressure-broadened by collisions in the atmosphere with hydrogen and helium. We calculate new $v \sin{i_p}$'s for each of these models and find that although the result from the $0.1\times$ pressure model is consistent at the $2.2\sigma$ level, the $10\times$ pressure model produces a $v \sin{i_p}$ that is $7.1\sigma$ lower (see Table \ref{tab:vsini_model_tests}). This difference is due to the fact that increasing the strength of pressure broadening creates broader features that require less rotational broadening to match the observed line widths. However, inferring rotational broadening with an atmospheric model that has $10\times$ pressure is unrealistic, so we can move past this discrepancy. Additionally, we note that although there are other atmospheric model grids, uncertainties related to model choice are insignificant compared to the radius uncertainty (Sec.~\ref{sec:spin_axis_inclination}) and the sky-plane uncertainty (Sec.~\ref{sec:obliquities}), which produce the dominant sources of error in the planet obliquity.

\begin{deluxetable}{lll}
\tablecaption{Model Tests and Resulting $v \sin{i_p}$'s}
\tablehead{
    \colhead{Model} & \colhead{ $v \sin{i_p}$ [\units{km~s^{-1}}] } & \colhead{$\sigma$--difference}
    }
\startdata 
    1200\units{K}, $\log{(g)}=4.5$ (fiducial) & $8.73 \pm 0.14$ \\
    1100\units{K}, $\log{(g)}=4.0$ & $8.78 \substack{+0.14 \\ -0.13}$ & $0.3\sigma$\tablenotemark{a} \\
    1100\units{K}, $\log{(g)}=5.0$ & $8.69 \pm 0.14$ & $0.2\sigma$ \\
    1300\units{K}, $\log{(g)}=4.0$ & $8.69 \pm 0.14$ & $0.2\sigma$ \\
    1300\units{K}, $\log{(g)}=5.0$ & $8.40 \substack{+0.15 \\ -0.16}$ & $1.6\sigma$ \\
    50\% Solar C/O & $8.71 \substack{+0.13 \\ -0.12}$ & $0.1\sigma$ \\
    150\% Solar C/O & $8.60 \pm 0.14$ & $0.7\sigma$ \\
    0.1x Pressure & $9.14 \pm 0.12$ & $2.2\sigma$ \\
    10x Pressure & $6.76 \pm 0.24$ & $7.1\sigma$ \\
\enddata
\tablenotetext{a}{Comparison to fiducial model}
\label{tab:vsini_model_tests}
\end{deluxetable}

\subsection{Measuring \texorpdfstring{$P_\textsubscript{rot}$}{Prot} for VHS~1256~b} \label{sec:rotation_period}

Most brown dwarfs exhibit low-amplitude variability ($\sim 0.2\%-2\%$) \citep{Metchev+2015}, and variability greater than 10\% is exceptionally rare \citep{Eriksson+2019}. The detection of rotational modulation can often be modelled using sinusoidal light curves in order to infer the rotation period \citep{Apai+2013, Vos+2017}. With this method, VHS~1256~b has a rich history of rotation period observations \citep{Bowler+2020b,Zhou+2020,Zhou+2022}. 

\citet{Zhou+2020} observed VHS~1256~b for the longest continuous coverage (36 hours), using the Spitzer Space Telescope/IRAC and finds a rotation period of $P_\textsubscript{rot}=22.04 \pm 0.05\units{hours}$ (hereafter, the Spitzer period). \citealt{Zhou+2022} then observed VHS~1256~b for 42 hours over four segments using the Hubble Space Telescope/WFC3, and finds complex light curves that can be explained by a combination of three sine waves corresponding to three periods: $18.8 \pm 0.2\units{hr}, 15.1 \pm 0.2\units{hr}$, and $10.6 \pm 0.1\units{hr}$ (hereafter, the HST periods). The discrepancy between the Spitzer period and HST periods likely arose from the fact that short time windows were used to sample a constantly evolving light curve, and the observed light curves do not fully encapsulate the evolution patterns, as described in Section 5.1 of \citet{Zhou+2022}. When the windows of continuous observation become long enough, as those presented in \citet{Apai+2021}, the period can be more robustly determined by the periodogram analysis.

To calculate the planet obliquity for VHS~1256~b, we use the Spitzer period over the HST periods for two reasons. First, the Spitzer observations have a much longer continuous coverage (36 hours) than the HST observations, and have a more precise constraint on the periodicity. Second, the Spitzer light curve is fully consistent with a single sine wave model, and the HST light curve is not. This suggests that the atmospheric evolution in VHS~1256~b is likely quieter during the Spitzer epoch, and thus the Spitzer period is less impacted by weather and atmospheric dynamics \citep{Zhou+2022}. In addition, we note that the 0.05 hour uncertainty is simply the result of error propagation in the least-squared fit, and does include systematic noise due to model limitations. Since this is small uncertainty is overly optimistic, we adopt a more conservative uncertainty of 10\% (2.2 hours) for our planet obliquity calculation. This wide range in uncertainty is supported by the analysis of fine structures in the periodogram of Luhman 16's TESS light curve (Section 4.3, \citealt{Apai+2021}).

\subsection{Measuring \texorpdfstring{$i_p$}{ip} for VHS~1256~b} \label{sec:spin_axis_inclination}

With the projected equatorial velocity ($v \sin{i_p}$) and rotation period ($P_\textsubscript{rot}$) in hand, the final ingredient required to calculate the spin axis inclination ($i_p$) of VHS~1256~b is the radius ($R$). We use the effective blackbody radius:

\be
R = \sqrt{\frac{L}{4\pi\sigma_b T_\text{eff}^4}},
\ee

\noindent where $L$ is the bolometric luminosity $\log{(L_\text{bol}/ \Lsun)} = -4.55 \pm 0.009$ \citep{Miles+2023}, $\sigma_b$ is the Stefan-Boltzmann constant, and $T_\text{eff}$ is the effective temperature $T_\text{eff} = 1200 \pm 100\units{K}$.\footnote{
    We choose this effective temperature to be consistent with our $v \sin{i_p}$ analysis in Sec.~\ref{sec:vsini}. We choose a conservative error estimate based on the grid-based atmospheric modelling done in \citet{Miles+2023}, which has intervals of $100\units{K}$.
}
This yields a blackbody radius $R = 1.20 \substack{+0.22 \\ -0.18}$ $\units{R_\text{Jup}}$. At first, one may presume to combine these quantities as follows\footnote{
    Specifically, this is the case when uncertainties are not negligible. This equation is correct in the limit where measurement errors approach zero \citep{Masuda+Winn2020}.
}:

\be
i_p = \arcsin{\left[\frac{P_{\text{rot}} \times v\sin{i_p}}{2\pi R}\right]}.
\label{eq:planet_spin_axis_inc}
\ee

However, this does not account for correlations between $v$ and $v \sin{i_p}$, so we follow the method described in \citet{Masuda+Winn2020} to infer the posterior of $i_p$ using two assumptions: 

1. The datasets \{$d_v,d_{v \sin{i_p}}$\} for $v=2\pi R/P_{\text{rot}}$ and $v \sin{i_p}$ are independent, so the likelihood for $v=2\pi R/P_{\text{rot}}$ and $v \sin{i_p}$ is separable.

2. The quantities $v$ and $i_p$ are independent, so the priors on $v$ and $i_p$ are separable. 

From these assumptions, the posterior PDF for $\cos{i_p}$ using Bayes' theorem is

\begin{align}
p(\cos{i_p} \mid D) \propto & \; \mathcal{P}_{\cos{i_p}}(\cos{i_p}) \int_0^{v_\text{break-up}} p(d_v \mid v) \nonumber \\
& \times p\left(d_{v \sin{i_p}} \;\middle|\; v\sqrt{1-\cos^2{i_p}}\right) \mathcal{P}_v(v)dv,
\label{eq:vsini_and_v}
\end{align}

\noindent where D is the whole dataset \{$d_v,d_{v \sin{i_p}}$\}, $\mathcal{P}_{\cos{i_p}}(\cos{i_p})$ is the prior on $\cos{i_p}$, which is uniform between 0 and 1, and $p(d_v \mid v)$ is the likelihood for $v$.
$p\left(d_{v \sin{i_p}} \;\middle|\; v\sqrt{1-\cos^2{i_p}}\right)$ is the likelihood for $v \sin{i_p}$ as calculated in Section \ref{sec:vsini}, and $\mathcal{P}_v(v)$ is the prior on $v$, which is uniform between 0 and the break-up velocity. 

\begin{figure}
    \centering
    \includegraphics[width=1.0\linewidth]{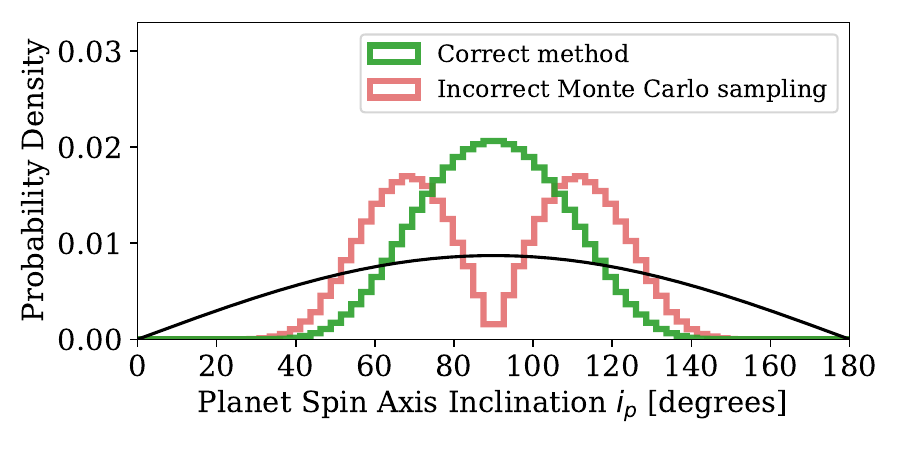}
    \caption{Posterior distribution of spin axis inclination $i_p$ for VHS~1256~b with the correct method (green, following equation~\ref{eq:vsini_and_v}, and with incorrect Monte Carlo sampling (red, following equation~\ref{eq:planet_spin_axis_inc}). The correct method yields $i_p = 90^\circ \pm 18^\circ$. These distributions are compared to a random inclination distribution (black), which is a uniform distribution in $\cos{i_p}$.
    }
    \label{fig:spin_axis_inc}
\end{figure}

Integrating equation~\eqref{eq:vsini_and_v} numerically, and converting this posterior PDF in $\cos{i_p}$ into samples of $i_p$ by rejection sampling yields the distribution shown in Figure~\ref{fig:spin_axis_inc}. Note that we also display the posterior PDF from incorrect Monte Carlo sampling (red) for $i_p$ by using equation~\eqref{eq:planet_spin_axis_inc} without accounting for correlations between $v$ and $v \sin{i_p}$. The discrepancy is particularly evident for VHS~1256~b since $v \sim v \sin{i_p}$, and therefore the spin axis orientation of VHS~1256~b is near perpendicular to our line of sight. The correct posterior peaks at $90^\circ$ and has a 68\% highest probability density interval (HDPI) of $[72^\circ, 108^\circ]$, or equivalently $90^\circ \pm 18^\circ$.

\subsection{Measuring \texorpdfstring{$i_o$}{io} for VHS~1256~b} \label{sec:orbital_inclination}

To measure the orbital inclination, we use the relative astrometry of VHS~1256~b (4 epochs over 6 years, \citealt{Dupuy+2023}) that traces out a small orbit arc. To fit this orbital motion, we use a Bayesian rejection sampling algorithm \citep{OFTI}, implemented using \texttt{orbitize!}\citep{orbitize}. In addition to relative astrometry, we use a system mass of $0.152 \pm 0.010 \units{\Msun}$ from Table 2 of \citet{Dupuy+2023} and a parallax of $47.27 \pm 0.47 \units{mas}$ from \citet{Gaia_EDR3}.

Our orbit fit\footnote{
    Here we present the median and 68\% highest density probability intervals, following \citet{Dupuy+2023}. The full posterior can be accessed upon request to the corresponding author.
}
yields a semimajor axis $a = 383 \substack{+99 \\ -150}\units{au}$, eccentricity $e = 0.70 \substack{+0.07 \\ -0.09}$, and inclination $i_o = 23 \substack{+10 \\ -13}\units{^\circ}$, which we will use to calculate the planet obliquity for VHS~1256~b. This result is consistent with \citet{Dupuy+2023}, who instead used a different package called \texttt{LOFTI\_GAIA} \citep{lofti_gaia} to find $i_o = 24 \substack{+10 \\ -15} \units{^\circ}$. \texttt{LOFTI\_GAIA} is similar to \texttt{orbitize!}, but assumes exactly linear motion in the astrometric data and combines all astrometric data points into a singular position and velocity. Therefore, we opt to use \texttt{orbitize!} in case the orbital motion is slightly nonlinear.

\subsubsection{Jointly fitting Astrometry with Companion Radial Velocity}
\label{sec:joint_fit}

To further constrain the orbital inclination measurement from \citet{Dupuy+2023}, we tested if obtaining a radial velocity (RV) measurement of VHS~1256~b relative to the system barycenter would be useful. This RV could be obtained from our MCMC analysis in Section \ref{sec:vsini}, in addition to further analysis to obtain the RV of the host binary. We estimate the RV upper limit of VHS~1256~b relative to the system barycenter as $\sim 1\units{km~s^{-1}}$ for a circular, edge-on orbit at $200\units{au}$. 

We use \texttt{orbitize!} to jointly fit the same astrometry as in Sec.~\ref{sec:orbital_inclination} along with simulated RV measurements (specifically, a small RV case: $0.0 \pm 0.1 \units{km~s^{-1}}$, and a large RV case: $0.5 \pm 0.1 \units{km~s^{-1}}$), with simulated RV errors motivated by RV constraints in the $v \sin{i_p}$ analysis in Sec.~\ref{sec:vsini}. With either simulated RV measurement, the new orbital inclination constraint is consistent with the original to $<0.2 \sigma$. Instead, an additional RV measurement is useful in constraining the argument of periapsis ($\omega$) and longitude of the ascending node ($\Omega$). This is because the RV provides information of the planet's direction in/out of the sky, and therefore orients the plane of the orbit. However, the $\omega$ and $\Omega$ posteriors are not covariant with other orbital elements (semi-major axis, eccentricity, and orbital inclination). 

\subsection{Measuring \texorpdfstring{$i_\text{AB}$}{iAB} for VHS~1256~AB} 
\label{sec:binary_inclination}

Unlike the companion VHS~1256~b, the binary host VHS~1256~AB has a well constrained orbit since its orbital period is much smaller. With a semi-major axis of $\sim 2\units{au}$, \citet{Dupuy+2023} observed VHS~1256~AB for more than half an orbit over a 6 year baseline with Keck/NIRC2. From these observations, they determine the line-of-sight inclination of the binary orbital plane to be $i_\text{AB} = 118.7^\circ \pm 1.0^\circ$.

\subsection{Measuring 3D Spin-Orbit Architecture of the VHS~1256-1257 System} \label{sec:obliquity}

In hand, we have constraints on the line-of-sight (LOS) inclinations of three angular momentum vectors in the VHS~system (see Fig.~\ref{fig:VHS_diagram}):

\begin{figure*}
    \centering
    \includegraphics[width=1.0\linewidth]{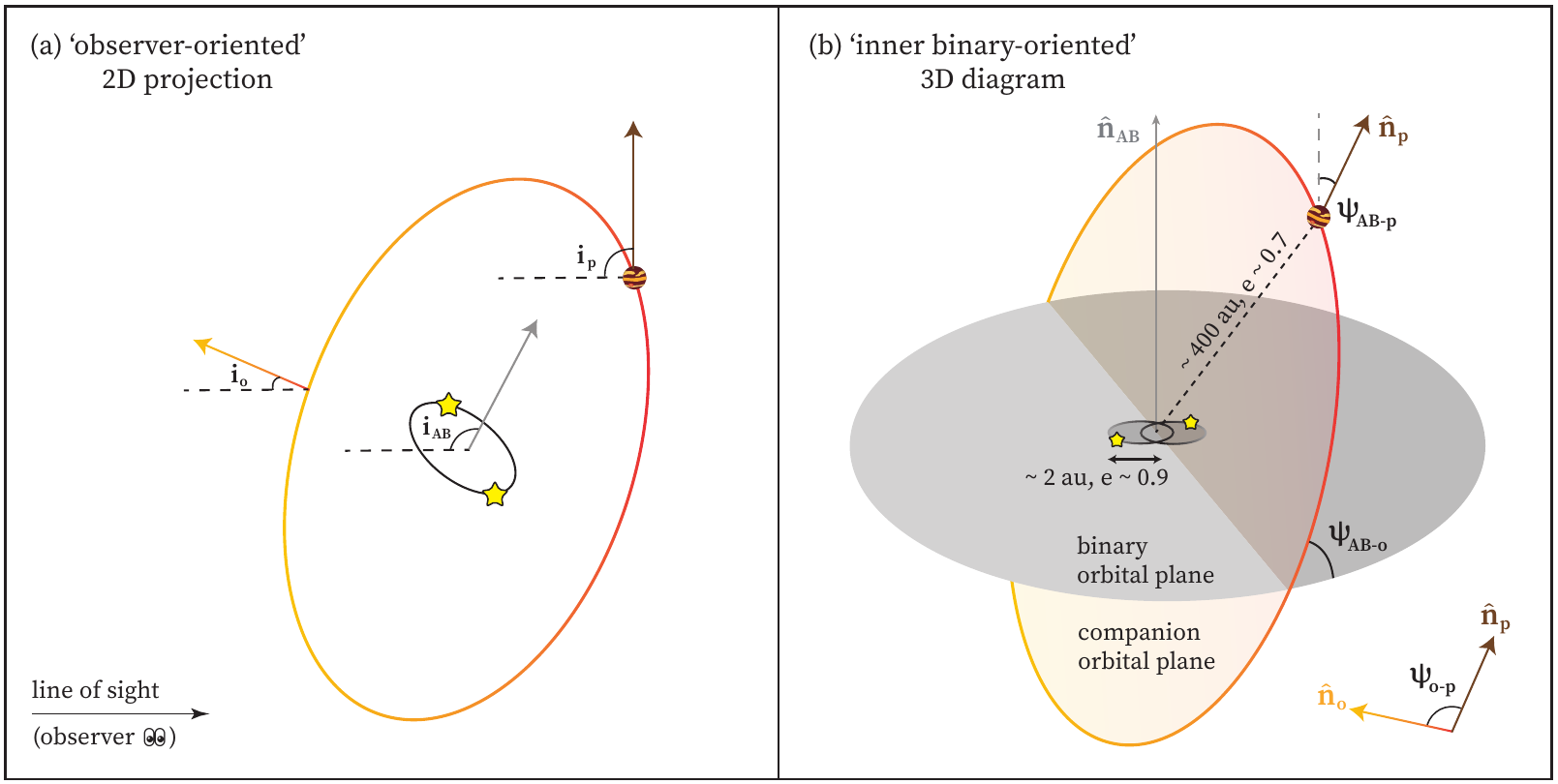}
    \caption{Schematic diagram of three angular momentum vectors in the VHS~system in 2D (left) and 3D (right). Left: line-of-sight inclinations for the spin axis of VHS~1256~b ($i_p$), the orbit normal of VHS~1256~b ($i_o$), and the orbit normal of the host binary VHS~1256~AB ($i_\text{AB}$) are shown relative to an observer on the left. Right: Here we highlight the true 3D mutual inclinations between angular momentum vectors. Specifically $\psi_\text{o-p}$ is the planet obliquity, and $\psi_\text{AB-p}$ ($\psi_\text{AB-o}$) is the mutual inclination between the binary orbital plane and planet spin axis (planet orbit normal). Note that all orbits are eccentric, and not to scale.
    }
    \label{fig:VHS_diagram}
\end{figure*}

1. $i_p$, the LOS inclination for the spin angular momentum of VHS~1256~b $\hat{n}_p$ (this work),

2. $i_o$, the LOS inclination for the orbital angular momentum of VHS~1256~b $\hat{n}_o$ (this work), and

3. $i_\text{AB}$, the LOS inclination for the orbital angular momentum of VHS~1256~AB $\hat{n}_{\text{AB}}$ \citep{Dupuy+2023}.

\begin{figure*}
    \centering
    \includegraphics[width=1.0\linewidth]{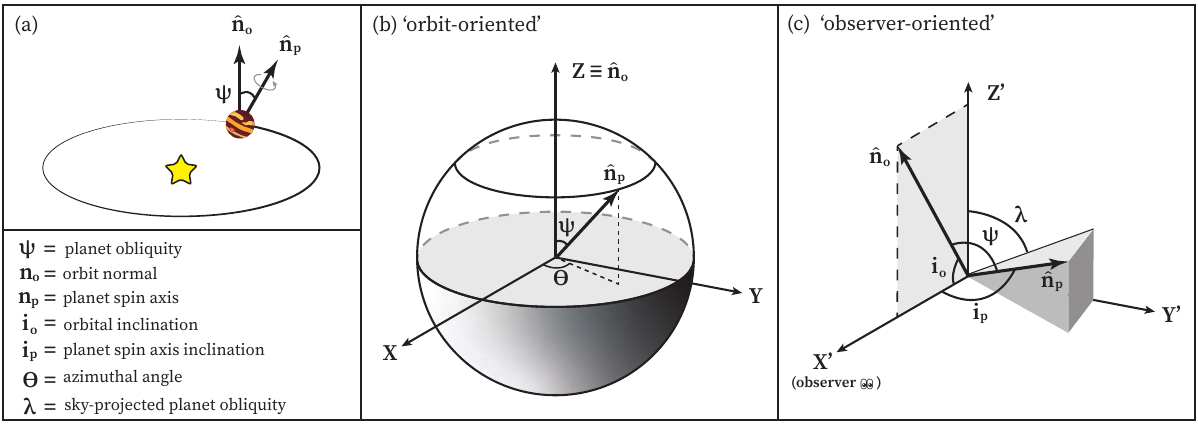}
    \caption{Schematic diagram of all relevant angles related to a planet obliquity. (a) Cartoon diagram showing that planet obliquity is the angle between a planet orbit normal and planet spin vector in 3D, (b) An orbit-oriented coordinate system with the orbit normal along the $Z-$axis. (c) An observer-oriented coordinate system with observer along the $X'-$axis, and the line of nodes along the $Y'-$axis. These two coordinate systems are related by a $90^\circ - i_o$ rotation along the $Y=Y'-$axis. These diagrams were inspired by \citealt{Fabrycky+Winn2009} and \citealt{Dong+DFM2023}.
    }
    \label{fig:coordinate_systems}
\end{figure*}

There are three 3D angles ($\psi_\text{o-p}, \psi_\text{AB-o}, \psi_\text{AB-p}$) between angular momentum vectors $\hat{n}_p$, $\hat{n}_o$, and $\hat{n}_{\text{AB}}$. $\psi_\text{o-p}$ is the true planet obliquity, which is the 3D angle between $\hat{n}_o$ and $\hat{n}_p$. Similarly, $\psi_\text{AB-o}$ is the true orbit-orbit mutual inclination. Lastly, $\psi_\text{AB-p}$ is 3D angle between the binary orbit $\hat{n}_{\text{AB}}$ and planet spin axis $\hat{n}_p$. To visualize these many angles, we encourage the reader to explore Figure~\ref{fig:VHS_diagram}, which orients the 3 angular momentum vectors in the system, and Figure~\ref{fig:coordinate_systems}, which illustrates the relevant coordinate systems.

We calculate these 3D angles in two ways, characterized by the coordinate system used: `observer-oriented' and `orbit-oriented' (Fig.~\ref{fig:coordinate_systems}). Only the observer-oriented method has been used to constrain all previous planet obliquities \citep{Bryan+2020, Bryan+2021, Palma-Bifani+2023}. With this method, the planet obliquity $\psi_\text{o-p}$ is given by:

\be
\cos{\psi_\text{o-p}} = \cos{i_o}\cos{i_p} + \sin{i_o}\sin{i_p}\cos{\lambda_\text{o-p}},
\label{eq:planet_obliquity} \\
\ee

\noindent where $\lambda_\text{o-p}=\Omega_o-\Omega_p$ is the sky-plane angle between the companion orbit and companion spin axis (see Fig.~\ref{fig:coordinate_systems}(c) for a visualization). For a detailed derivation of equation~\eqref{eq:planet_obliquity}, we refer the reader to \citet{Fabrycky+Winn2009} and \citet{Dong+DFM2023}. 

Similarly, we can calculate the 3D mutual inclinations $\psi_\text{AB-o}$ and $\psi_\text{AB-p}$ as:

\begin{align}
\cos{\psi_\text{AB-o}} &=\cos{i_\text{AB}}\cos{i_o} + \sin{i_\text{AB}}\sin{i_o}\cos{\lambda_\text{AB-o}},
\label{eq:orbit-orbit_inc} \\
\cos{\psi_\text{AB-p}} &= \cos{i_\text{AB}}\cos{i_p} + \sin{i_\text{AB}}\sin{i_p}\cos{\lambda_\text{AB-p}},
\label{eq:binary_orbit-spin_inc}
\end{align}

\noindent where $\lambda_\text{AB-o}=\Omega_\text{AB}-\Omega_o$, and $\lambda_\text{AB-p}=\Omega_\text{AB}-\Omega_p$. $\Omega_\text{AB}$ defines the sky-plane orientation of the binary orbit. We have constraints on $\Omega_o$ and $\Omega_\text{AB}$ from orbit fits, but $\Omega_p$ is currently not observable. 

In this observer-oriented method, the typical assumption for the sky-plane angle $\lambda_\text{o-p}$ is uniform between 0 and $2\pi$ (or equivalently 0 to $\pi$), which corresponds to a prior for $\psi_\text{o-p}$ that is randomly oriented. However, if a system truly had zero planet obliquity, assuming $\lambda_\text{o-p}$ to be uniform would bias $\psi_\text{o-p}$ to larger values.

We introduce an orbit-oriented method, that does not involve $\lambda_\text{o-p}$, but instead allows for an explicit prior for $\psi_\text{o-p}$. The advantage of this method is that having a flexible prior for $\psi_\text{o-p}$ can come from physical theories. If we choose the prior $\psi_\text{o-p}$ to be randomly oriented for a single planet obliquity measurement, this new method should yield equivalent results.

We construct this orbit-oriented method using Bayes' Theorem, and seek to find the posterior\footnote{
    Here, we use $\psi = \psi_\text{o-p}$ for brevity, but this also holds for $\psi_\text{AB-o}$ and $\psi_\text{AB-p}$.
}
$p(\psi|i_p, i_o) \propto \mathcal{P}(\psi)p(i_p|\psi, i_o)$. Here, $\mathcal{P}(\psi)$ is easily interchangeable and we can choose $\mathcal{P}(\psi)=\frac{1}{2}\sin{\psi}$ as an explicit prior, motivated by $\hat{n}_o$ and $\hat{n}_p$ being uncorrelated (or equivalently, randomly oriented). 

To construct the likelihood $p(i_p|\psi, i_o)$, we move to an orbit-oriented coordinate system as shown in Figure~\ref{fig:coordinate_systems}(b), where $\theta$ is the azimuthal angle of the planet spin axis. Following the coordinate transform from observer-oriented to orbit-oriented as detailed in \citet{Dong+DFM2023}, it can be shown that:

\be
\cos{i_p} = \cos{\psi}\cos{i_o} + \sin{\psi}\sin{i_o}\cos{\theta}.
\ee

Subsequently, we follow the variable transform described in Appendix A of \citet{Campante+2016}, which leads to:

\be
\begin{split}
p\big(i_p \big| \psi, i_o\big) &= p\big(\theta \big| i_o,\psi\big) \bigg|\frac{\partial \theta}{\partial i_p} \bigg| = \frac{1}{\pi} \bigg|\frac{\partial \theta}{\partial i_p} \bigg| \\
&= \frac{1}{\pi} \frac{\sin{i_p}}{\sqrt{\sin^2{\psi}\sin^2{i_o} - (\cos{\psi}\cos{i_o} - \cos{i_p})^2}},
\end{split}
\ee

\noindent or zero if the argument of the square root is negative. With the orbit-oriented method, one does not need to worry about histogram bin sizes or interpolating a discrete CDF since the posterior $p(\psi|i_p, i_o)$ can be obtained directly.

\subsubsection{Planet Obliquity Constraints} \label{sec:obliquities}

\begin{figure}
    \centering
    \includegraphics[width=1.0\linewidth]{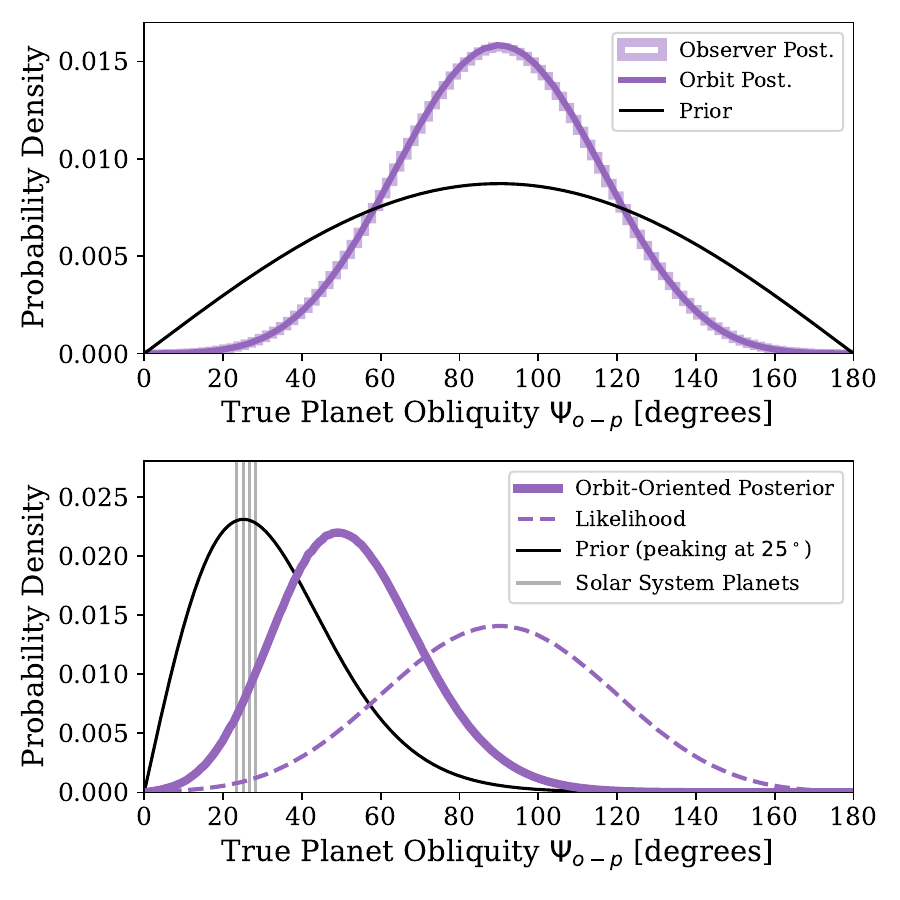}
    \caption{Top panel: Normalized posterior distribution for the planet obliquity $\psi_\text{o-p}$, using a randomly oriented prior (black), which is uniform in $\cos{\psi}$. The observer-oriented and orbit-oriented methods show agreement. Bottom panel: Same as top panel, but with a different prior (black). This prior is a Fisher distribution that peaks at $25^\circ$, which is chosen to roughly match Solar System planet obliquities (from small to large: Earth, Mars, Saturn, Neptune). Mercury, Venus, Jupiter and Uranus are excluded as they have planet obliquities near $0^\circ, 90^\circ$, and $180^\circ$, preventing a simple unimodal prior. In subsequent analysis, we choose to use the posterior from the upper panel.
    }
    \label{fig:obliquity_comparison}
\end{figure}

\begin{figure}
    \centering
    \includegraphics[width=1.0\linewidth]{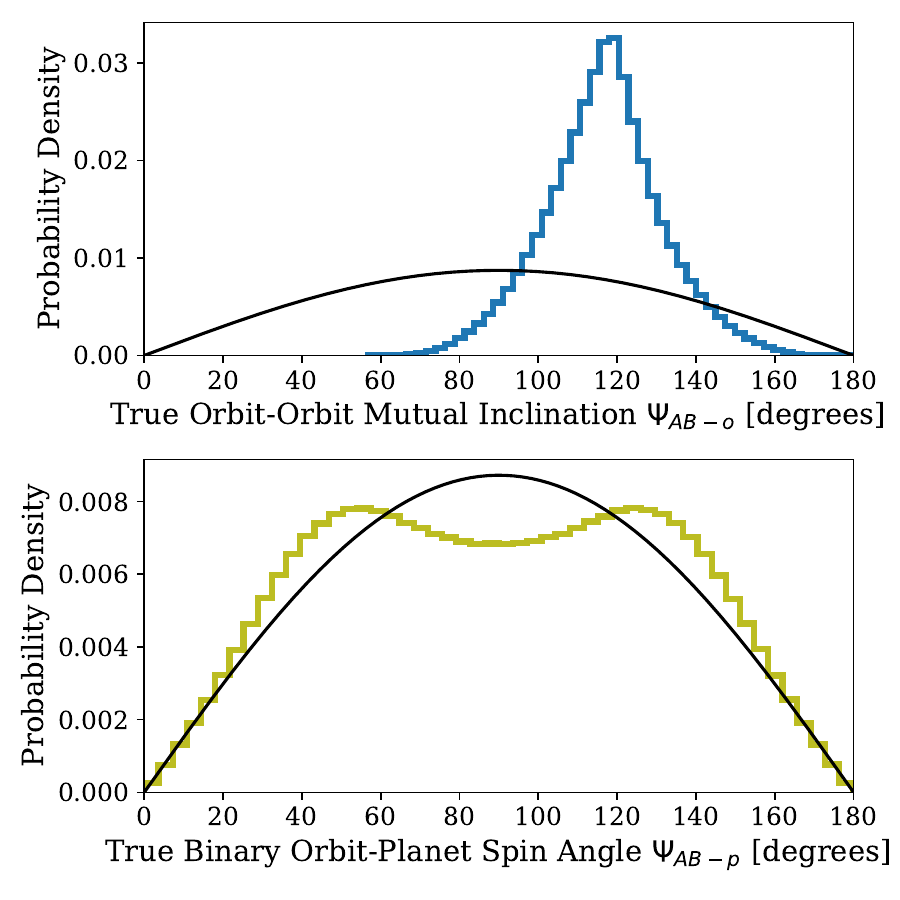}
    \caption{Top panel: Normalized posterior distribution for the orbit-orbit mutual inclination $\psi_\text{AB-o}$. Bottom panel: Normalized posterior distribution for the 3D angle between the binary orbit and planet spin axis $\psi_\text{AB-p}$. These distributions are compared to a randomly oriented prior (black), which is uniform in $\cos{\psi}$.
    }
    \label{fig:mutual_3D_inc}
\end{figure}

Using the observer-oriented and orbit-oriented methods, we calculate the planet obliquity $\psi_\text{o-p}$ posterior for VHS~1256~b in Figure~\ref{fig:obliquity_comparison} (top), compared to a prior where $\hat{n}_o$ and $\hat{n}_p$ are uncorrelated. Since the assumptions in both methods are the same ($\mathcal{P}(\lambda)$ is uniform when $\mathcal{P}(\psi) \propto \sin{\psi}$), the posteriors for both methods are the same. As a comparison, we investigate how the posterior changes with a Solar-system-like prior, where planet obliquities tend to be small. For this, we use a Fisher distribution that peaks at 25 degrees as a prior. This posterior is shown in Figure~\ref{fig:obliquity_comparison} (bottom), which is heavily influenced by the prior. It seems more likely that VHS~1256~b formed differently that the Solar System, and it is more likely part of an ensemble of systems that has randomly oriented planet obliquities. Therefore, we opt to use the posterior from the upper panel of Figure~\ref{fig:obliquity_comparison} in subsequent analysis. Additionally, we show the posterior for $\psi_\text{AB-o}$ and $\psi_\text{AB-p}$ in Figure~\ref{fig:mutual_3D_inc}.

\begin{figure}
    \centering
    \includegraphics[width=1.0\linewidth]{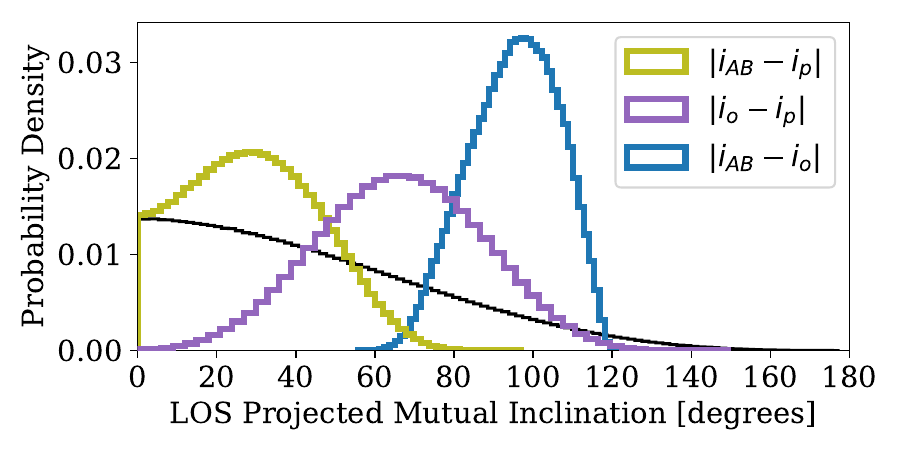}
    \caption{Posterior distributions for the line-of-sight projected mutual inclinations. These are lower limits on the true 3D mutual inclinations $\psi_\text{o-p}, \psi_\text{AB-o}$ and $\psi_\text{AB-p}$. These distributions are compared to a randomly oriented projected inclination distribution (black), where $i_o$, $i_p$, and $i_\text{AB}$ have all been drawn from uniform distributions in $\cos{i}$.
    }
    \label{fig:projected_obliq}
\end{figure}

\begin{deluxetable}{lll}
\tablecaption{Measured Parameters}
\tablehead{
    \colhead{Parameter} & \colhead{Measured Value} & \colhead{Ref.}
    }
\startdata 
    $v \sin{i_p}$ & $8.7 \pm 0.1~\units{km~s^{-1}}$ & This work\\
    $P_\textsubscript{rot}$ & $22.04 \pm 0.05~\units{hr}$ & \citet{Zhou+2020}\\
    $i_p$ & $90^\circ \pm 18^\circ$ & This work\\
    $i_o$ & $23 \substack{+10 \\ -13}^\circ$ & This work\\
    $i_o$ & $24 \substack{+10 \\ -15}^\circ$ & \citet{Dupuy+2023}\\
    $i_\text{AB}$ & $118.7 \pm 1.0^\circ$ & \citet{Dupuy+2023}\\
    $|i_o - i_p|$ & $67\substack{+21 \\ -22}^\circ$ & This work\\
    $|i_\text{AB} - i_o|$ & $97^\circ \pm 12^\circ$ & This work\\
    $|i_\text{AB} - i_p|$ & $29 \substack{+16 \\ -20}^\circ$ & This work\\
    $\psi_\text{o-p}$ & $90^\circ \pm 25^\circ$ & This work\\
    $\psi_\text{AB-o}$ & $118\substack{+12 \\ -16}^\circ$ & This work\\
    $\psi_\text{AB-o}$ & $115^\circ \pm 14^\circ$ & \citet{Dupuy+2023}\\
    $\psi_\text{AB-p}$ & $55\substack{+30 \\ -16}^\circ$ or $125\substack{+16 \\ -30}^\circ$ & This work\\
\enddata
\tablecomments{
    The three inclinations $i_p, i_o, i_\text{AB}$ presented here are along our line-of-sight. The line-of-sight mutual inclinations $|i_o - i_p|, |i_\text{AB} - i_o|$, and $ |i_\text{AB} - i_p|$ are lower limits on the true deprojected angles $\psi_\text{o-p}, \psi_\text{AB-o}$, and $ \psi_\text{AB-p}$. For the line-of-sight mutual inclinations and true deprojected angles, we quote the mode and 68\% highest density probability intervals.}
\label{tab:parameters}
\end{deluxetable}

The planet obliquity $\psi_\text{o-p}$ mode and 68\% HDPI is $90^\circ \pm 25^\circ$. The true orbit-orbit mutual inclination $\psi_\text{AB-o}$ mode and 68\% HDPI is $118\substack{+12 \\ -16}^\circ$. Since the 3D angle between the binary orbit and planet spin axis $\psi_\text{AB-p}$ is bimodal, we calculate the mode and 68\% HDPI for each half of the distribution below and above $90^\circ$ as $55\substack{+30 \\ -16}^\circ$ or $125\substack{+16 \\ -30}^\circ$. These results are summarized in Table~\ref{tab:parameters}. $\psi_\text{o-p}$ and $\psi_\text{AB-o}$ show evidence of significant misalignment, whereas $\psi_\text{AB-p}$ looks similar to a randomly oriented prior. $\psi_\text{AB-o}$ provides the best constraint, because the sky-plane angle $\lambda_\text{AB-o}$ is well determined. On the other hand, $\psi_\text{o-p}$ and $\psi_\text{AB-p}$ have broad constraints since the sky-plane orientation of the planet spin axis $\Omega_p$ is not observable. However, $\psi_\text{o-p}$ is better constrained than $\psi_\text{AB-p}$ since $i_o$ and $i_p$ share little overlap, whereas $i_\text{AB}$ falls within $i_p$.

Moreover, we can calculate the lower limit on the planet obliquity $\psi_\text{o-p}$ and true mutual inclinations $\psi_\text{AB-o}$ and $\psi_\text{AB-p}$ \citep{Bowler+2017}:

\begin{align}
\psi_\text{o-p} &\geqslant |i_o - i_p|,
\label{eq:proj_planet_obliquity} \\
\psi_\text{AB-o} &\geqslant |i_\text{AB} - i_o|,
\label{eq:proj_orbit-orbit_inc} \\
\psi_\text{AB-p} &\geqslant |i_\text{AB} - i_p|.
\label{eq:proj_binary_orbit-spin_inc}
\end{align}

We calculate the mode and 68\% HDPI of $|i_o - i_p|, |i_\text{AB} - i_o|$, and $|i_\text{AB} - i_p|$ as $67\substack{+21 \\ -22}^\circ$, $97^\circ \pm 12^\circ$, and $29 \substack{+16 \\ -20}^\circ$ respectively (Fig.~\ref{fig:projected_obliq}, Table~\ref{tab:parameters}). We notice that $|i_o - i_p|$ and $|i_\text{AB} - i_o|$ show strong evidence for misalignment.

To quantify this misalignment, we calculate the probability that each $\psi$ distribution falls into an `aligned' state, which we define as $\psi \in (0^\circ, 20^\circ)$, or a `misaligned' state, which we define as $\psi \in (20^\circ, 180^\circ)$. We calculate the Bayesian odds ratio $p(\text{misaligned}|\text{data})/p(\text{aligned}|\text{data})$, where the numerator is the integral of the posterior distribution $p(\psi|\text{data})$ from $20^\circ - 180^\circ$, and the denominator is the integral of the posterior distribution $p(\psi|\text{data})$ from $0^\circ - 20^\circ$.

As a reference, we calculate the odds ratio for a randomly oriented prior which is uniform in $\cos{\psi}$. In the absence of any data (e.g. the prior shown by the black curve in Figure~\ref{fig:mutual_3D_inc}), the odds ratio favoring misalignment to alignment is 32:1 ($2.2\sigma$)\footnote{
    To calculate this $2.2\sigma$ significance, we use the 68-95-99.7 rule: $y=1-\erf{\Big(\frac{x}{\sqrt{2}}\Big)}$, where $y=\Big(\frac{1}{1+32}\Big)$ is the chance of alignment and therefore $x=2.2$ is the significance.
    }.
The odds ratio for the planet obliquity $\psi_\text{o-p}$ is more significant than the prior at 1061:1 ($3.3\sigma$). The odds ratio for $\psi_\text{AB-o}$ is extremely significant at $>10^8:1$ ($>5.7\sigma$). The odds ratio for $\psi_\text{AB-p}$ is similar to random orientation, at 33:1 ($2.2\sigma$). To visualize, we are seeing a planetary-mass companion with an edge-on spin-axis on a near face-on orbit around a binary-host with a near edge-on orbit.

We have found that the VHS~1256 system has evidence of misalignment among all angular momentum vectors. With these blueprints in hand, we are ready to examine which theories can possibly explain the unique spin-orbit architecture and unusual system properties in VHS~1256.

\section{Discussion: Possible Formation Histories} \label{sec:dicussion}

VHS~1256 is a rare, very low-mass triple system (see the bias-corrected multiplicity fraction in Fig. 1 of \citealt{Offner+2023}). The setup is hierarchical, with an inner equal-mass ratio binary VHS~1256~AB (M$_{\rm tot}$$\sim 0.14\units{\Msun}$, $a_{\rm inner} \sim 2\units{au}$), and a planetary-mass tertiary ($\sim 12-16$M$_{\rm Jup}$, $a_{\rm outer} \sim 400\units{au}$) \citep{Dupuy+2023}. Both inner and outer orbits are eccentric, $e \sim 0.9, 0.7$ respectively. In this work, we find that all three observationally constrained angular momentum vectors in this system, namely the tertiary orbit normal, the binary orbit normal, and the tertiary spin vector, are all misaligned relative to each other. Now we consider a range of planet and star formation scenarios which could explain these misaligments and other measured system properties.

In Section \ref{sec:planet-like}, we consider a starting point where VHS~1256~b formed like a planet. In this core accretion scenario, the companion would form bottom up in the circumstellar disk, with spin and orbital angular momenta comparable to that of the disk. Considering close-in planets forming around close binaries, observations of circumbinary planets show that planetary and stellar orbits are nearly coplanar $\lesssim 3^\circ$ \citep{Li+2016}. Taken together, if VHS~1256~b formed by core accretion, our expectation is that the planet spin axis, planet orbit normal, and binary orbit normal should be primordially aligned. This is inconsistent with the system's present misaligned state. So something else would be needed to produce these observed misalignments.

In Section \ref{sec:star-like}, we consider an alternative -- and considerably more promising -- starting point where VHS~1256~b formed by gravitational collapse. This process is fast and turbulent, and therefore should produce angular momentum vectors with broadly random orientations \citep{Offner+2016, Lee+2019}. 

\subsection{Planet-like Formation} \label{sec:planet-like}

If bottom up formation produces aligned planet obliquities, some process must produce the observed misalignments afterwards. In our Solar System, one of these processes may be violent collisions, attributed to the near-$90^\circ$ tilt of Uranus \citep{Kegerreis+2018, Rogoszinski+Hamilton2021}. Could this account for the misalignments we see in the VHS~1256 system?

Considering the likely outcomes of gravitational interactions between VHS~1256~b and another object, a collision is more likely if the escape velocity from VHS~1256~b $v_\text{esc,p}$ is smaller than the escape velocity from the host star at the semi-major axis of VHS~1256~b $v_\text{esc,$\star$}$. If the opposite is true, then ejection rather than collision is favored. For the VHS~1256 system, this ratio of escape velocities is:

\be
\frac{ v_\text{esc,p} }{ v_\text{esc,$\star$} } = \left(\frac{ M_p / R_p }{ M_\star / a_p } \right)^{1/2} \sim 240 \gg 1,
\ee

\noindent where $M_p = 12~M_\text{Jup}$, $R_p = 1.2~R_\text{Jup}$, $a_p = 400\units{au}$, and $M_\star = 0.14~\Msun$. Given the large ratio, it is much more likely that a nearby object would be presently ejected rather than collide. 

We can also consider whether VHS~1256~b's -$90^\circ$ obliquity could arise through secular spin-orbit resonances, a mechanism potentially responsible for Saturn's $27^\circ$ obliquity and anticipated for at least close-in exoplanetary systems  \citep{Ward+Hamilton2004,Millholland+Batygin2019, Millholland+Laughlin2019, Su+Lai2020, Li2021}. A spin-orbit resonance occurs when the spin axis precession rate matches the nodal precession rate. This is unlikely to occur in the VHS~1256 system due to the wide orbital separation of VHS~1256~b -- this predicts a prohibitively high spin axis precession timescale of $\sim 10^{13} \units{years}$, older than the age of the universe.

Perhaps outer Kozai-Lidov oscillations \citep{Farago+Laskar2010, Naoz+2017} could be relevant. Here, the binary torque from VHS~1256~AB could have `kicked' VHS~1256~b onto an inclined orbit, leading to a nonzero obliquity. However, like spin-orbit resonances, the relevant timescale for this effect is too long ($\sim 10^{10} \units{years}$) due to the wide separation of VHS~1256~b.

Another way to potentially tilt the orbit of VHS~1256~b and produce a misaligned obliquity is with a stellar flyby \citep{Rodet+2021, Rodet+Lai2022}. We estimate how close an expected flyby can be \citep{Adams2010} within the age of the system ($\sim 140\units{Myr}$, \citealt{Dupuy+2023}):

\begin{align}
\Gamma^{-1}  &= \left( n\sigma \overline{v} \right)^{-1} \nonumber\\
             &= 140 \left( \frac{0.14/{\rm pc}^3}{n} \right) 
                    \left( \frac{5100\units{au}}{b} \right)^2
                    \left( \frac{26\units{km~s^{-1}}}{\overline{v}} \right) \units{Myr},
\label{eq:flyby_time}
\end{align}

\noindent where $n$ is the stellar density, and $\overline{v}$ is the velocity dispersion, which we choose based on the stellar environment of the solar neighbourhood \citep{Brown+Rein2022}. $b\sim 5100\units{au}$ is the closest approach of an expected flyby. Using an impulse approximation \citep{Rickman1976} for solar-mass stellar flyby, we find that this would cause a minimal change in velocity for VHS~1256~b ($\Delta v\sim 0.1\%~v_{\rm esc}$). Stellar flybys are also unlikely to explain any of the observed misalignments.

In sum, none of these mechanisms could misalign an initially aligned VHS~1256 system due in large part to the wide separation of VHS~1256~b. These mechanisms aside, we note that a fatal flaw of proposing to form VHS~1256 b bottom up like a planet is the fact that it is a wide-separation massive planet around a brown dwarf binary. The circumbinary disk would not have enough material to form a planet that massive that far out \citep{Armitage2020}. We set the planet formation scenario aside to move onto a more promising avenue.

\subsection{Star-like Formation} \label{sec:star-like}

Generally, objects forming top-down form 1) faster; 2) often in multiple systems, and 3) with a broader distribution of separations and eccentricities \citep{Offner+2023,Heggie1975}. To zeroth order, this seems like a good fit for the VHS~1256~system. 

A recent review breaks this picture down and describes three broad mechanisms for multiple star formation:  disk fragmentation, fragmentation of a core or filament, and dynamical interactions \citep{Offner+2023}. 

Disk fragmentation typically occurs in a massive disk at $\sim 10 - 500\units{au}$ \citep{Kratter+Lodato2016,Bate2018, Jennings+Chiang2021,Offner+2023}. The low mass of the host binary (0.141 M$_{\odot}$), high companion-to-stellar mass ratio ($M_p/M_\star \sim 0.1$), and wide separation of the companion compared to potential disk radii present challenges to this formation pathway \citep{Kratter+Lodato2016}. However, uncertainties associated with disk evolution and other processes leave enough room for disk fragmentation to remain a possible mechanism for the formation of VHS~1256~b.

Core and filament fragmentation produce binaries and higher order multiples, possibly through turbulent fragmentation \citep{Bate+2002, Bate2009, Bate2012, Chabrier+2014}. Initially these are expected to have wide separations and spin axes that are randomly oriented \citep{Offner+2016, Tokovinin2017,Lee+2019}. All three misalignments in the VHS~1256 system can be accounted for. 

Let us add in the other measured system properties like mass, mass ratio, separation, and eccentricity to this picture. Lower mass binaries tend to form with more equal masses, which the central VHS~1256~AB binary follows \citep{Bate2012,Fontanive+2018}. The large eccentricity of the tertiary VHS~1256 b is characteristic of brown dwarf companions (e.g. \citealt{Bowler+2020a}). One option is VHS~1256 b could have formed as an isolated object and been subsequently captured \citep{Clarke+Pringle1991, Bate2018}. In this scenario, enough energy has to be dissipated in this encounter for VHS~1256~b to become gravitationally bound. However, neither the low-mass binary nor the likely low-mass circumbinary disk would be effective energy sinks. Without this, capture is unlikely. 

Perhaps this triple system formed top-down together. Substellar triple systems are uncommon, but possible. While simulations of cloud fragmentation show that multiple systems initially form at wide separations, then gas-dynamical friction \citep{Lee+2019, Offner+2023} or dynamical interactions post-fragmentation can cause orbital decay \citep{Bate2012}. This progression would be needed to account for the close binary separation of VHS~1256~AB ($\sim 2\units{au}$). 

Taking all the observations together, we propose the following as a promising formation scenario for the VHS~1256 system. Initially, core/filament fragmentation formed a higher order (i.e. triple or quadruple) gravitationally-bound system. The presence of gas then causes a pair to decay to smaller separations via dynamical friction, forming VHS~1256~AB. An alternative and less likely pathway is that through dynamical interactions post-fragmentation, the lowest mass member of the quadruple gets ejected. During this ejection, angular momentum and energy leave the system, allowing the orbit of VHS~1256~AB to harden and shrink to its present day close separation (e.g. \citealt{Reipurth+Mikkola2015}). This story can account for the misalignments of the binary orbit, companion orbit, and companion spin axis, as well as the low masses, high mass ratios, large eccentricities, and two orders of magnitude difference in separation between the central binary and the tertiary.  

\section{Conclusions} \label{sec:conclusions}

In this study, we constrain a key piece in the angular momentum architecture of the VHS~1256 system: the line-of-sight spin-axis inclination of the planetary-mass companion VHS~1256~b. We find that VHS~1256~b spins edge-on, with $i_{\rm p} = 90^\circ \pm 18^\circ$. This, in combination with VHS~1256~b travelling on a near face-on orbit, implies a highly misaligned planet obliquity of $\psi_\text{o-p} = 90^\circ \pm 25^\circ$. Due to the chance orientation of this spin axis and orbit normal, we have provided the first exoplanetary obliquity that is not bimodal. 

This obliquity is reminiscent of the $\sim 98^\circ$ planet obliquity that Uranus maintains. However, it is unlikely that VHS~1256~b formed in a planet-like manner for several reasons. The typical timescale for planet-like scenarios to excite a nonzero obliquity, such as spin-orbit resonances, scale quickly to restrictively long timescales with companion distance. In addition to evidence for a nonzero obliquity, our constraints show that the orbital plane of the binary host VHS~1256~AB is misaligned with \textit{both} the spin axis and orbit normal of its companion. Specifically, we find a 3D orbit-orbit mutual inclination $\psi_\text{AB-o}$ mode and 68\% HDPI of $118\substack{+12 \\ -16}^\circ$. Since the 3D angle between the binary orbit and planet spin axis $\psi_\text{AB-p}$ is bimodal, we calculate the mode and 68\% HDPI for each half of the distribution below and above $90^\circ$ as $55\substack{+30 \\ -16}^\circ$ or $125\substack{+16 \\ -30}^\circ$.

We quantified this misalignment by calculating the probability that each $\psi$ distribution falls into an `aligned' state, which we define as $\psi \in (0^\circ, 20^\circ)$, or a `misaligned' state, which we define as $\psi \in (20^\circ, 180^\circ)$. The odds ratio for $\psi_\text{o-p}$, $\psi_\text{AB-o}$ , and $\psi_\text{AB-p}$ is 1061:1 ($3.3\sigma$), $>10^8:1$ ($>5.7\sigma$), and 33:1 ($2.2\sigma$), respectively. We are seeing a planetary-mass companion with an edge-on spin-axis on a near face-on orbit around a binary-host with a near edge-on orbit.

We propose that top-down formation, through core/filament fragmentation, followed by gas-driven migration or ejection is the most promising scenario to explain how the VHS~1256 system formed. These scenarios have been illustrated in turbulent magneto-hydrodynamic \citep{Lee+2019} and large scale radiation hydrodynamical simulations \citep{Bate2012}, and reasonably explains all of the observed misalignments and measured properties in the VHS~1256 system.

To date, there are four planet obliquity measurements (2M0122-2439~b, \citealt{Bryan+2020}; HD~106906~b, \citealt{Bryan+2021}; AB~Pic~b, \citealt{Palma-Bifani+2023}; and VHS~1256~b, this work), and further efforts to increase this sample will be crucial to obtain statistically significant population-level inferences. With the James Webb Space Telescope now online, we will have greater access to more planetary-mass companion rotation periods, which is the most challenging observable to obtain. Soon, we will have access to bona fide exoplanetary obliquities for planetary-mass companions that are truly exoplanets. This is particularly exciting as it would be a direct comparison to envision how planets in our Solar System fit into the broader extrasolar context.

\begin{acknowledgments}
We thank Sarah Blunt, Eugene Chiang, Samantha Fassnacht, Sam Hadden, Max Moe, Dang Pham, Joshua Speagle, Ethen Sun, and Scott Tremaine for useful discussions.

M.L.B. acknowledges support by NSERC, the Heising-Simons Foundation, and by the Connaught New Researcher Award from the University of Toronto. This work used the Immersion Grating Infrared Spectrometer (IGRINS) that was developed under a collaboration between the University of Texas at Austin and the Korea Astronomy and Space Science Institute (KASI) with the financial support of the Mt. Cuba Astronomical Foundation, of the US National Science Foundation under grants AST-1229522 and AST-1702267, of the McDonald Observatory of the University of Texas at Austin, of the Korean GMT Project of KASI, and Gemini Observatory.
\end{acknowledgments}

\bibliography{main}{}
\bibliographystyle{aasjournal}

\end{document}